\documentclass[twocolumn,amsmath,amssymb,floatfix,superscriptaddress,prl,noeprint,aps]{revtex4-2}
\usepackage{bm,graphicx,url,epsf,color}
\usepackage{amssymb}
\usepackage{amsthm}
\usepackage[dvipsnames]{xcolor}
\usepackage{overpic}
\usepackage{graphicx}

\usepackage{verbatim}
\usepackage{physics}
\usepackage{overpic}
\usepackage[colorlinks=true,linkcolor=blue,citecolor=blue]{hyperref}

\usepackage[utf8]{inputenc}
\usepackage{amsmath}
\usepackage{amsfonts}
\usepackage{bm}
\usepackage{mdframed}
\usepackage[normalem]{ulem}
\usepackage{siunitx}

\newcommand{\Rb}{$^{87}$Rb }

\newcommand{\stkout}[1]{\ifmmode\text{\sout{\ensuremath{#1}}}\else\sout{#1}\fi}

\newcommand{\spinu}{{\color{BrickRed}\uparrow}}
\newcommand{\spind}{{\color{Blue}\downarrow}}

\date{\today}

\begin{document}
\title{Local vs non-local dynamics in cavity-coupled Rydberg atom arrays}
\author{Zeno Bacciconi}
\affiliation{International School for Advanced Studies (SISSA), Via Bonomea 265, I-34136 Trieste, Italy}
\author{Hernan B. Xavier}
\affiliation{International School for Advanced Studies (SISSA), Via Bonomea 265, I-34136 Trieste, Italy}
\affiliation{The Abdus Salam International Centre for Theoretical Physics (ICTP), Strada Costiera 11, 34151 Trieste, Italy}
\author{Matteo Marinelli}
\affiliation{University of Trieste, Physics Department, Via A. Valerio 2, 34127 Trieste, Italy}
\author{Devendra Singh Bhakuni}
\affiliation{The Abdus Salam International Centre for Theoretical Physics (ICTP), Strada Costiera 11, 34151 Trieste, Italy}
\author{Marcello Dalmonte}
\affiliation{The Abdus Salam International Centre for Theoretical Physics (ICTP), Strada Costiera 11, 34151 Trieste, Italy}

\begin{abstract}
Locality is a transversal principle that governs quantum dynamics of many-body systems. However, for cavity embedded systems, such fundamental notion is hindered by the presence of non-local cavity modes, leaving space for new possible dynamical behaviors. Here, we investigate the real-time dynamics of low-energy excitations in one dimensional Rydberg atom arrays coupled to a global cavity mode. We derive an effective description in terms of a Tavis-Cummings-Ising model,  whose phase diagram features ordered and disordered phases. The non-local nature of the cavity mode drastically affects the emergent meson and string dynamics. Mesons hybridize coherently with the cavity photons, leading to composite meson-polaritons excitations. Strings, differently from local interacting theories, acquire a finite kinetic energy thanks to non-local cavity-mediated interactions between the underlying domain-walls. We then conclude by presenting a new concrete experimental blueprint for a cavity QED Rydberg atom array simulator where the physics outlined in this work can be realized.
\end{abstract}
\maketitle

\paragraph{Introduction. -}There is presently considerable interest in exploring the real time dynamics of strongly interacting quantum matter in quantum simulators and computers~\cite{buluta2009quantum,cirac2012goals}. A major goal of this endeavor is to shed light on the role of low-lying excitations, in particular, in the context of gauge theories and spin models, where basic constituents such as mesons and strings play a prominent role~\cite{zohar2015quantum,dalmonte2016lattice,banuls2020simulating,bauer2023quantum,kormos2017real,GeZi_Nori_prb2023_supercond,GeZi_Nori_prl2024_scars}. Remarkable recent experiments have begun to investigate this within the realm of Abelian gauge theories as well as Ising-like models, exploring string ~\cite{martinez2016real,bernien2017probing,gonzalez2024observation,liu2024string,de2024observation,cochran2024visualizing} as well as meson-driven dynamics~\cite{liang2024observationanomalousinformationscrambling,farrell2024quantum} and false-vacuum decay~\cite{lagnesePasquale_prb2021_falsevacuum,zhu2024probing}. The dynamics considered so far has been driven by local couplings, which severely constrain the effective interactions between quasiparticles.
That is not generally the case for cavity embedded quantum matter \cite{rmp2013_ritsch_esslinger_cavityatom,Mivehvar2021,Schlawin_2022apr_cavityrev,Bloch_2022nat_cavityrev,GarciaVidal_2021science_cavityrev}, an emerging class of quantum many body systems characterized by the interplay between local degrees of freedom (atoms \cite{Lonard_nature2017,nie2021dissipative,Chanda_quantum2021,Chiriacò_prb2023_lightmatter,Riccardo_2023prr_cavityQED,zhu2024nonreciprocal,chang_marino_prr2024_spinglass,chang_marino_prr2024_fieldtheory}, spin \cite{schiro2012phase,viehmann2013observing,dalmonte2015realizing,Gelhausen_2016SciPost,Li_2020prr,Li_2021pra,Macri_prl2024_rydbergmesons,Chiocchetta_natcomm2021,Pupillo_prb2024_multifractality,mallick2024stringbreakingdynamicsising}, electrons \cite{JarcFausti_2023,AppuglieseFaist_science2022,EnknerGraziotto_prx2024_vonklitzing,enknerfaist_2024_fqhenhanced,bacciconi2024theoryfractionalquantumhall,PhongCiuti_prl2023_cherncavity,HagenmullerPupillo_prl2017_chargetransport,bacciconi2023topological}) and non-local cavity modes --- leading to a plethora of phenomena not accessible in traditional spin models. It is thus an open question whether such couplings to global modes can give rise to qualitatively new meson and string dynamics, and whether that can be experimentally probed. 

\begin{figure*}
\centering

\raisebox{0.1\height}{\begin{overpic}
    [width=0.25\linewidth]{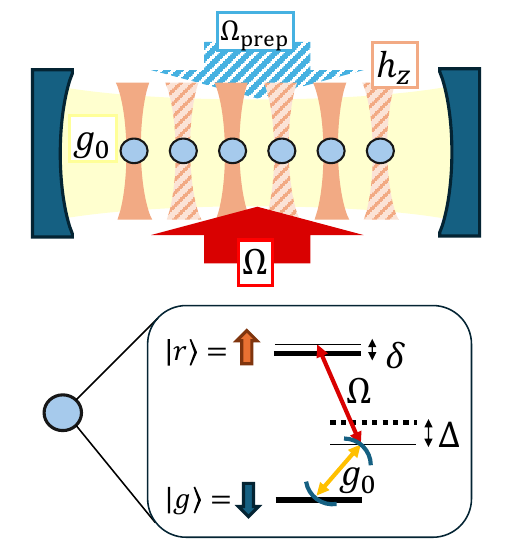}
    \put(0,0){(a)}
\end{overpic}}
\hspace{0.01\linewidth}
\begin{overpic}
    [width=0.3\linewidth]{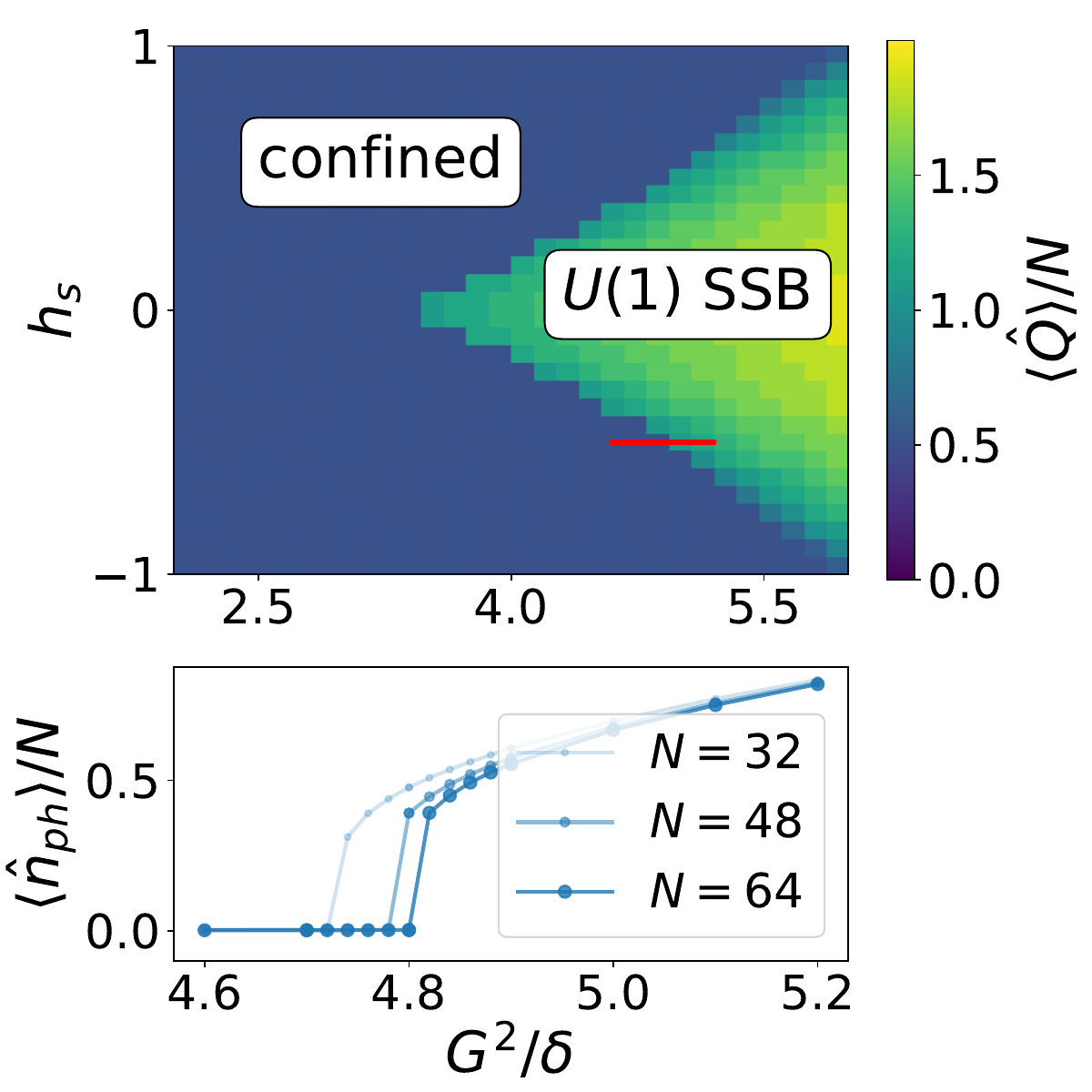}
    \put(16,30){(c)}
    \put(16,50){\textcolor{white}{(b)}}

\end{overpic}
\hspace{0.02\linewidth}
  \raisebox{0.1\height}{
  \begin{overpic}
    [width=0.35\linewidth]{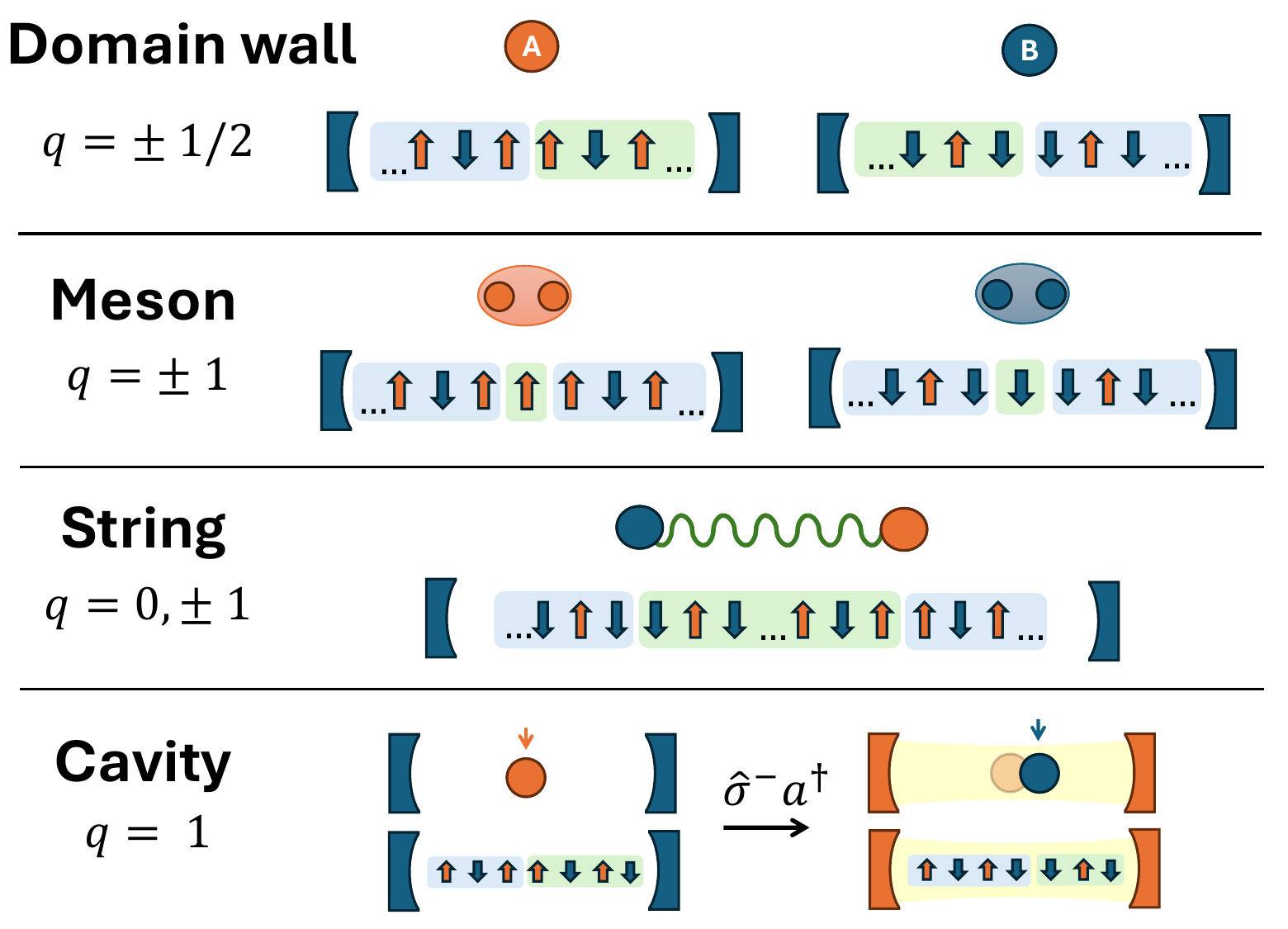}
    \put(-3,0){(d)}  
\end{overpic}}
    \caption{(a) Sketch of the set-up under consideration, an array of individually controlled rydberg atoms coupled to a cavity mode via a two-photon transition assisted by an external laser $\Omega$, giving a tunable coupling $g=\Omega g_0/\Delta$. (b-c) Phase diagram of the TC-Ising model as a function of $h_z$ and $G=g\sqrt{N}$ via (b) the ground state charge sector $Q$ (ED and $N=10$) and (c) photon number (DMRG); the latter at a fixed $h_z=-0.5$ and varying $G$ (red line of (b)). In both (b) and (c) $V=\delta=1.$ and $\lambda=0.$ (d) Sketch of the relevant low-energy degrees of freedom and their $U(1)$ charge in the confined antiferromagnetic phase. }
    \label{fig:sketch}
\end{figure*}

In this work, we consider a setup consisting of an array of neutral atoms, whose ground state is coupled to a Rydberg state via a two-photon transition involving a cavity field, see Fig \ref{fig:sketch}(a). Our proposal builds upon and combines outstanding achievements in the field of cavity QED \cite{Lonard_nature2017,yan_2023_Stamper_superradiant,ho2024_Stamper_optomechanicalselforganizationmesoscopicatom} and Rydberg atom experiments \cite{bernien2017probing,Browaeys2020,BluvsteinLukin_2023nature}, with the promise of achieving unexplored regimes where both local and non-local interactions are dominating over cavity losses and realize coherent light-matter many body dynamics.

Within this setting, locally interacting spins couple to a global mode in a controllable manner, realizing a perfect setting where to investigate hybrid meson and string dynamics. After deriving the phase diagram of the model, we show that the cavity field has a drastic impact on its low-energy dynamics: different types of solitons emerge, featuring very distinct energetics due to cavity-assisted processes. Some of the solitons can propagate freely thanks to the cavity, while others are dynamically blocked, and effectively stay confined due to energetics.

The effect of light onto composite objects is drastic. Meson quasiparticles hybridize with light, generating a meson-polariton excitation that immediately de-localizes single mesons. 
In deconfined regimes instead, mesons melt during the real-time dynamics via a multi-stage local process involving a combination of spin and cavity excitations. String dynamics is instead informative about the fate of confinement in the presence of light. Remarkably we observe that strings, independently of their length, always acquire a finite kinetic energy, contrasting with local spin models where this is typically exponentially suppressed. This is due to cavity-mediated long-range interactions solitons, allowing for an exchange between string tension and kinetic energy. We illustrate this new class of dynamical behavior using exact (ED) and matrix product states (MPS)\cite{White1992,Schollwck2011,White_prb2020_globalsubspaceexpansion,HalatiKollath_prl2020_mpscavity,Bacciconi_2023scipost} simulations, and conclude our work by detailing a realistic experimental proposal. 

\paragraph*{Model Hamiltonian. -}We consider a collection of $N$ Rydberg atoms coupled to single cavity mode via a two photon transition through a far detuned intermediate state (see Fig.~\ref{fig:sketch}). This can be mapped onto a Tavis-Cumming model with an antiferromagnetic Ising interaction, in abbreviation Tavis-Cumming-Ising or TC-Ising. The Hamiltonian is~\cite{supmat}:
\begin{align}\label{eq:H_tot}
    &\hat{H}= \delta \hat{a}^\dagger \hat{a} +  h_z\sum_{j}(-1)^j \hat{\sigma}^z_j + V\sum_{j} \hat{\sigma}^z_j\hat{\sigma}^z_{j+1}\nonumber \\& + g\sum_j\left(\hat{a}\hat{\sigma}^+_j + \hat{a}^\dagger \hat{\sigma}_j^-\right)+ \lambda\; \hat{a}^\dagger \hat{a}\sum_{j}\hat{\sigma}_j^z,
\end{align}
where $\delta$ is the cavity detuning, $h_z$ a staggered longitudinal field achieved by AC Stark shifts coming from the optical tweezers, $V$ the Rydberg-Rydberg interaction truncated to the nearest neighbor for simplicity \footnote{Minor effects coming from longer-range interactions are discussed in \cite{supmat}.}, $g=g_0\Omega/\Delta$ the effective cavity-atom coupling strength and $\lambda=g_0^2/\Delta$ a cavity dependent Stark shift. A further laser $\Omega_{\mathrm{prep}}$ is needed for state preparation purposes. We set $\lambda=0$ for simplicity and $V=1$.

Importantly the model has a $U(1)$ symmetry associated to the number of photons and Rydberg excitations:
\begin{equation}
    \hat{Q}=\hat{a}^\dagger \hat{a} + \sum_{j}\hat{\sigma}^{+}_j \hat{\sigma}^{-}_j .
\end{equation}

In Fig.~\ref{fig:sketch}(b,c) we show the phase diagram obtained with exact diagonalization (ED) and matrix product state (MPS) simulations as a function of the \textit{collective} light-matter interaction $G=g\sqrt{N}$ and the longitudinal staggered field $h_z$ with fixed interaction $V=1$. Two phases are present (see Fig.~\ref{fig:sketch}(c) ), distinguished by the charge sector $Q$ in which the ground state lives. The ground state is antiferromagnetic when $Q=N/2$, corresponding to the confined phase\footnote{While $Q$ does not stricly reveal confinement, this can be detected by expectation values of string operators, in this case $W(i,j)=\prod_{i\leq l\leq j}\hat{\sigma}_l^x$ which we indeed checked to be vanishing in the confined phase.}.  When the charge $Q$ instead changes we find an in-plane ferromagnet with spontaneous symmetry breaking (SSB) of the continuous $U(1)$ symmetry related to the charge $Q$ \cite{Li_2021pra,GeZi_Nori_prr2024_superradiant,Minganti_Nori_iop2021_dpt,Minganti_Nori_prr2024_liouvillian}. In Fig.~\ref{fig:sketch}(c) we show the photon occupation for large systems obtained with DMRG, highlighting the first order nature of the transition \cite{supmat}. The phase diagram is similar to that of the longitudinal field Ising model, whose AFM phase can also be understood as a confined phase of a $Z_2$ lattice gauge theory. The strong coupling phase however is here enriched by the $U(1)$ SSB instead of being a simple disordered paramagnet. Note that also Dicke-Ising models \cite{Kai_prr2020_dickeIsing,Macri_prl2024_rydbergmesons} show the same feature with the SSB being of a discrete $Z_2$ instead of a continuous $U(1)$ symmetry.

In the following we will focus only on the AFM, or confined, phases dynamical properties by studying the low energy excitations depicted in Fig \ref{fig:sketch}(d). Note also that as long as the cost of creating a pair of domain walls $4V$ is larger than the light-matter collective coupling $g\sqrt{N}$, their number is going to be approximately conserved.

\paragraph*{Domain-wall propagation -} The presence of a global $U(1)$ symmetry induces strong differences between domain walls formed by up-up or down-down spins, unlike the standard phenomenology of $Z_2$ symmetric Ising models. We thus introduce two domain wall densities:
\begin{align}
    \hat{D}^A_j = |\spinu_j\spinu_{j+1}\rangle\langle\spinu_j\spinu_{j+1}|
    \quad  
    \hat{D}^B_j =|\spind_j\spind_{j+1}\rangle\langle\spind_j\spind_{j+1}|
\end{align}
which measure the presence of domain walls of type A or B at the bond between atom $j$ and $j+1$  ($j=0,\dots,N-1$). We study quenches of the light-matter interaction $g$, starting from initial classical domain wall states (Fig. \ref{fig:sketch}(d) ) with zero cavity photons. We remark that the system has no quantum dynamics in absence of $g$. 

\begin{figure}
    \centering
    \begin{overpic}
    [width=\linewidth]{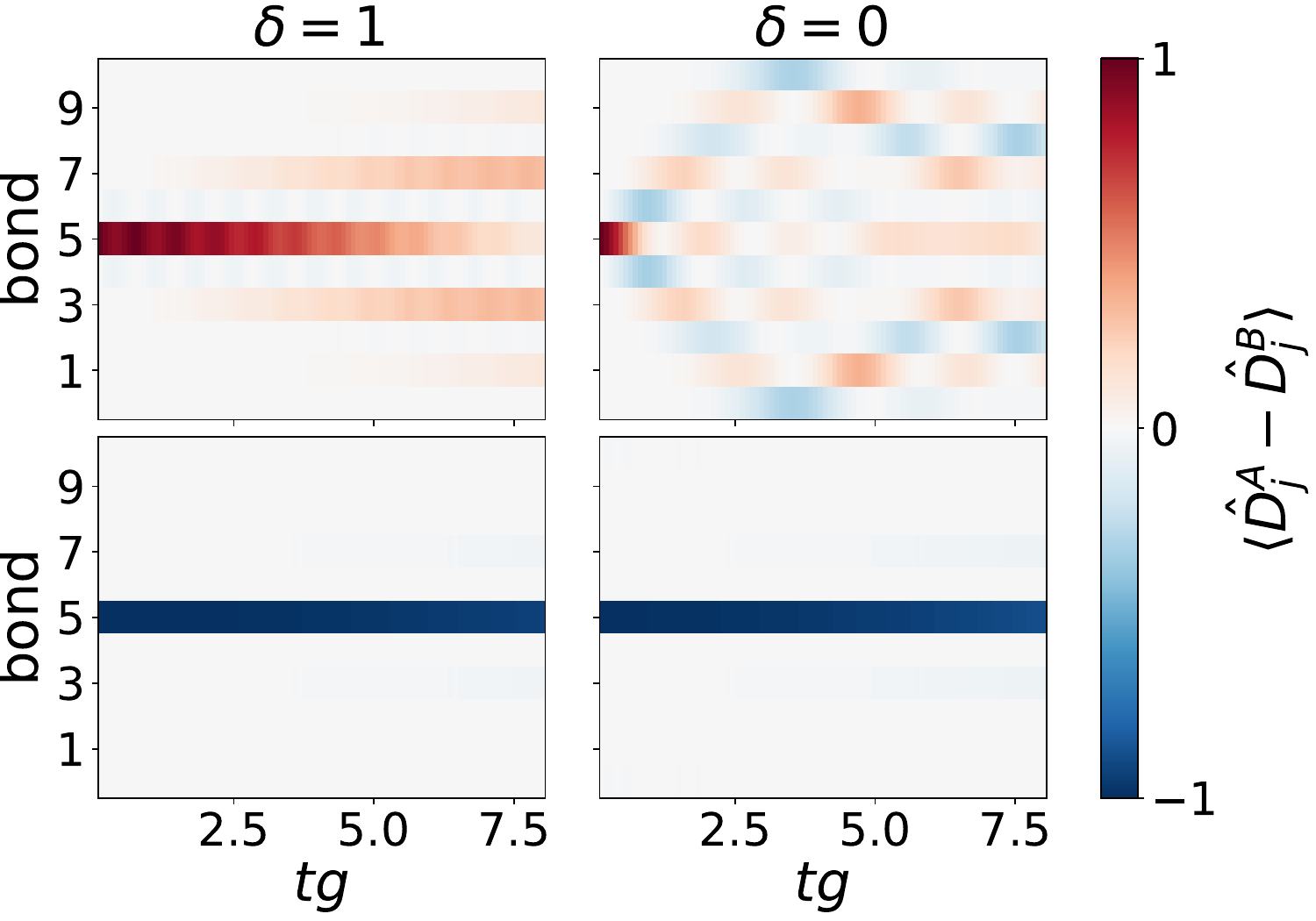}
    \put(9,11){(c)}
    \put(9,40){(a)}
    \put(48,11){(d)}
    \put(48,40){(b)}
        \end{overpic}

    \caption{Single domain wall cavity-mediated dynamics via domain wall density $\langle\hat{D}_j^A -\hat{D}_j^B\rangle$ obtained with ED. The system is initialized in a classical state with a single domain wall either of (a,b) type A or type B (c,d). The cavity photon energy is $\delta=1$ on the left (a,c) while $\delta=0$ on the right (b,d). $N=12$, $V=1$, $g= 0.12$ and $h_z=0$.}
    \label{fig:1dw_dynamics}
\end{figure}
In Fig. \ref{fig:1dw_dynamics} we show ED results for the propagation of type A (top row) and type B (bottom row) domain walls in two different regimes for the cavity detuning ($\delta= 1$ and $\delta=0$). The timescale of the cavity-induced dynamics is different in all four resulting cases but can be understood by simple perturbation theory. Starting from the cavity detuned regime, $\delta=1$, for domain walls of type $A$ we have a second-order process:
\begin{align}
\ket{\spinu\spind\spinu\spinu\spind\spinu}\ket{0}\;\rightarrow \;\ket{\spinu\spind\spinu\overline{\spind}\spind\spinu}\ket{1}\;\rightarrow \;\ket{\spinu\spind\spinu\overline{\spind\spinu}\spinu}\ket{0},
\end{align}
with a rate $J_A =g^2/\delta$. The peculiarity of this process is that the domain wall can only jump by two sites and will remain type $A$, as indeed clear in Fig. \ref{fig:1dw_dynamics}(a). Type $B$ domain walls instead can only move through creation and destruction of other domain walls, thus costing much more energy. At lowest order:
\begin{align}
\ket{\spind\spinu\spind\spind\spinu\spind}\ket{0}\;\rightarrow\; \ket{\spind\spinu\spind\spind\overline{\spind}\spind}\ket{1}\;\rightarrow \;\ket{\spind\spinu\spind\overline{\spinu\spind}\spind}\ket{0},
\end{align}
with a rate $J_B=g^2/(\delta + 4V)$ which effectively result in an apparent freezing of the domain wall.

Moving to small photon frequencies $\delta<g$ the propagation mechanism changes. The main difference is that, instead of being virtually populated by second-order processes, now the cavity can be resonantly populated. For example starting from a domain wall of type $A$ we have:
\begin{equation}
   \ket{\spinu\spind\spinu\spinu\spind\spinu}\ket{0}\;\rightarrow \;\ket{\spinu\spind\spinu\overline{\spind}\spind\spinu}\ket{1}
\end{equation}
happening with a rate $g$. Here the domain wall type can oscillate between $A$ and $B$ during the propagation as a photon is coherently exchanged back and forth. The other case of an initial condition with a domain wall of type $B$ (and no initial photons) does not allow these resonant processes but only the second order processes thus still giving a frozen dynamics.

\begin{figure}
    \centering
    \begin{overpic}
        [width=\linewidth]{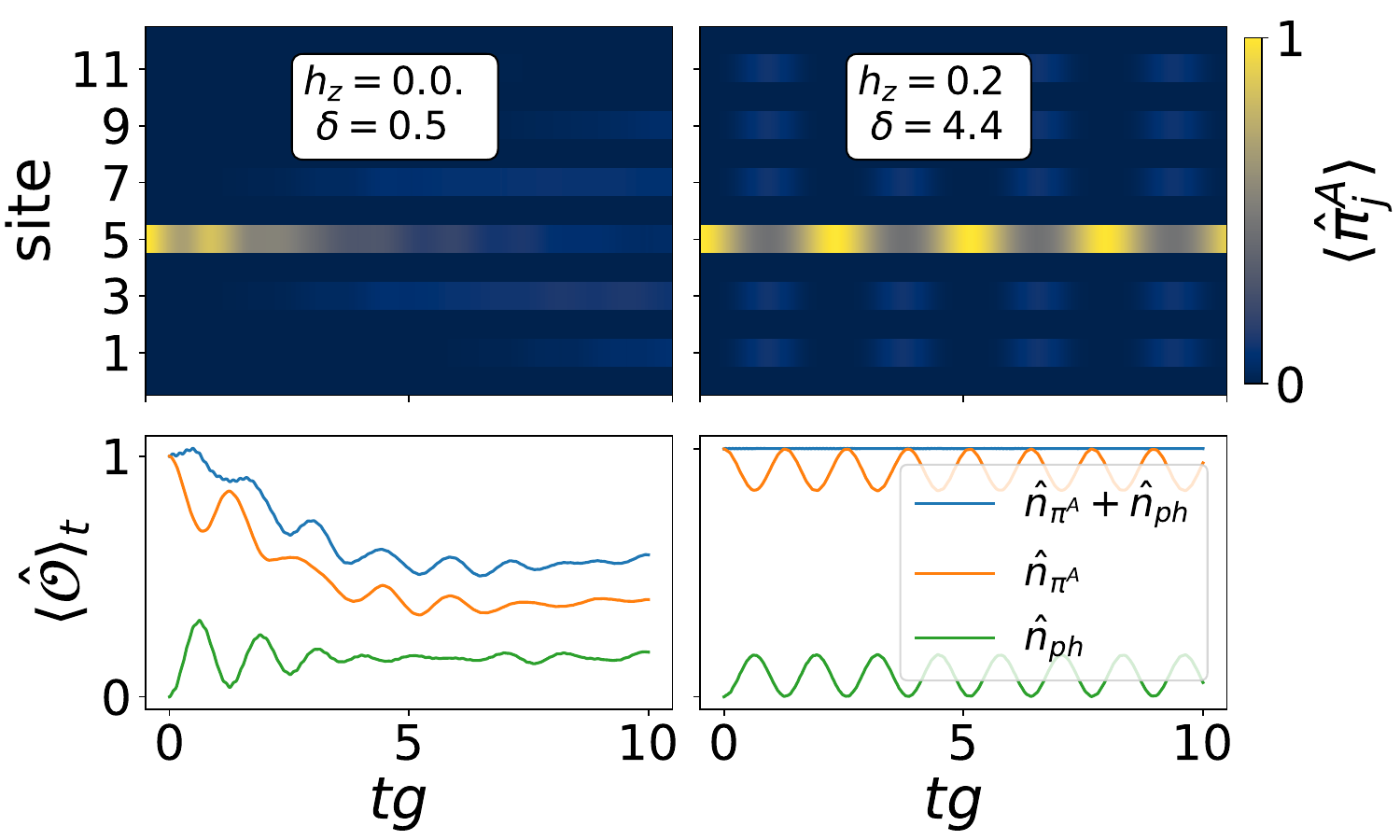}
    \put(11,33){\textcolor{white}{(a)}}
    \put(80,33){\textcolor{white}{(b)}}
    \put(43,24){(c)}
    \put(51,16){(d)}

    \end{overpic}

    \caption{Meson-polariton dynamics (a,c) in a  de-confined case ($h_z=0$ , $\delta=0.5$) and (b,d) in a confined case ($h_z=0.2$ , $\delta=4.4$) obtained with ED. Top panels (a,b) show the dynamics in real space via the meson density, while bottom panels (c,d) show the evolution in time of key observables (total meson number, photon number and their sum).  $V=1$, $g=0.1$ and $N=13$.}
    \label{fig:meson}
\end{figure}

\paragraph*{Meson-polaritons -} Another important low-energy excitation which can be built on top of an AFM ground state is a single spin-flip, i.e., a magnon or, in connection with the LGT interpretation, a meson. As for domain walls, we can identify two different type of mesons $A$ and $B$ corresponding respectively to a spin flipped up ($\spind$ to $\spinu$) and vice versa. In particular at $g=0$ the simplest meson states are classical configurations:
\begin{align}\label{eq:meson_state}
    &|\pi^{A/B}_j \rangle\ket{0} = \hat{\sigma}^{\pm}_j|\textit{AFM}\rangle\ket{0} \qquad \text{(for $j$ odd/even)}
\end{align}
where $|\textit{AFM}\rangle$ is the Néel state with up (down) spins on even (odd) sublattices and $\ket{0}$ the cavity vacuum, as sketched in Fig.\ref{fig:sketch}(d). In order to track the dynamics of these excitations we use the classical, type resolved, meson density:
\begin{align}
    \hat{\pi}_j^{A}= |\spinu_{j-1}\spinu_{j}\spinu_{j+1}\rangle \langle \spinu_{j-1}\spinu_{j}\spinu_{j+1}|\;,\nonumber\\
       \hat{\pi}_j^{B}= |\spind_{j-1}\spind_{j}\spind_{j+1}\rangle \langle \spind_{j-1}\spind_{j}\spind_{j+1}|\;.
\end{align} 
In Fig. \ref{fig:meson} we study the meson dynamics \footnote{Note that the classical meson density is a good indication of the presence of a meson only when this is close to its classical configuration, i.e. when $\delta \gg g$} starting from a classical meson state (Eq.~\ref{eq:meson_state}) in different regimes for $h_z$ and $\delta$, keeping $g=0.1$. The leftmost panels (a,c) show the dynamics in absence of confinement $h_z=0$. Here the two domain walls forming the meson split-up and propagate independently, as highlighted by a quickly decaying meson number $\hat{n}_{\pi^A}=\sum_j  \hat{\pi}_j^{A}$. In the right panels (b,d), we show the emergence of confinement with a finite $h_z=0.2$ for a cavity detuning $\delta=4V+2h_z=4.4$ which makes cavity photons $(\delta)$ and mesons $(4V+2h_z)$ resonant. Note that, despite confinement, only the combined sum (blue) of number of photons (green) and meson (orange) is conserved. This hybridization can be traced back to the formation of collective meson-polariton states which imprint a non-local dynamics on the initially local meson excitation, i.e., a de-localization of the meson on the whole spin-chain. In order to better understand the physics at play, we explicitly construct the collective states responsible for the meson-polariton formation:
\begin{align}
    &|C_1\rangle = |AFM\rangle |1\rangle,\;\;\;|C_2\rangle = \frac{1}{\sqrt{N_{odd}}}\sum_{j\,odd}|\pi^A_{j}\rangle |0\rangle.
\end{align}
The dynamics within this subspace then reduces to a simple two-level system dynamics:
\begin{align}
  \langle C_i|  \hat{H} |C_j\rangle = \begin{pmatrix}
      \delta &  g\sqrt{N_{odd}} \\
       g\sqrt{N_{odd}} & 4V +2 h_z
  \end{pmatrix} + const.
\end{align}
Kinetic terms for the mesons are order $g^4$ and thus neglected in this simple picture. Still from the two state model dynamics we can readily understand three important properties. (i) The non-local meson-polariton oscillations are collective, controlled by $G=g \sqrt{N}$, while the local dynamics is controlled by $g$. (ii) At resonance, $\delta=4V+2h_z$ there is a stronger hybridization giving rise to coherent Rabi oscillations. (iii) A local meson has a vanishing overlap $1/\sqrt{N_{odd}}$ with the collective states, thus limiting the amplitude of the Rabi oscillations.
\begin{figure}
    \centering
    \begin{overpic}
        [width=\linewidth]{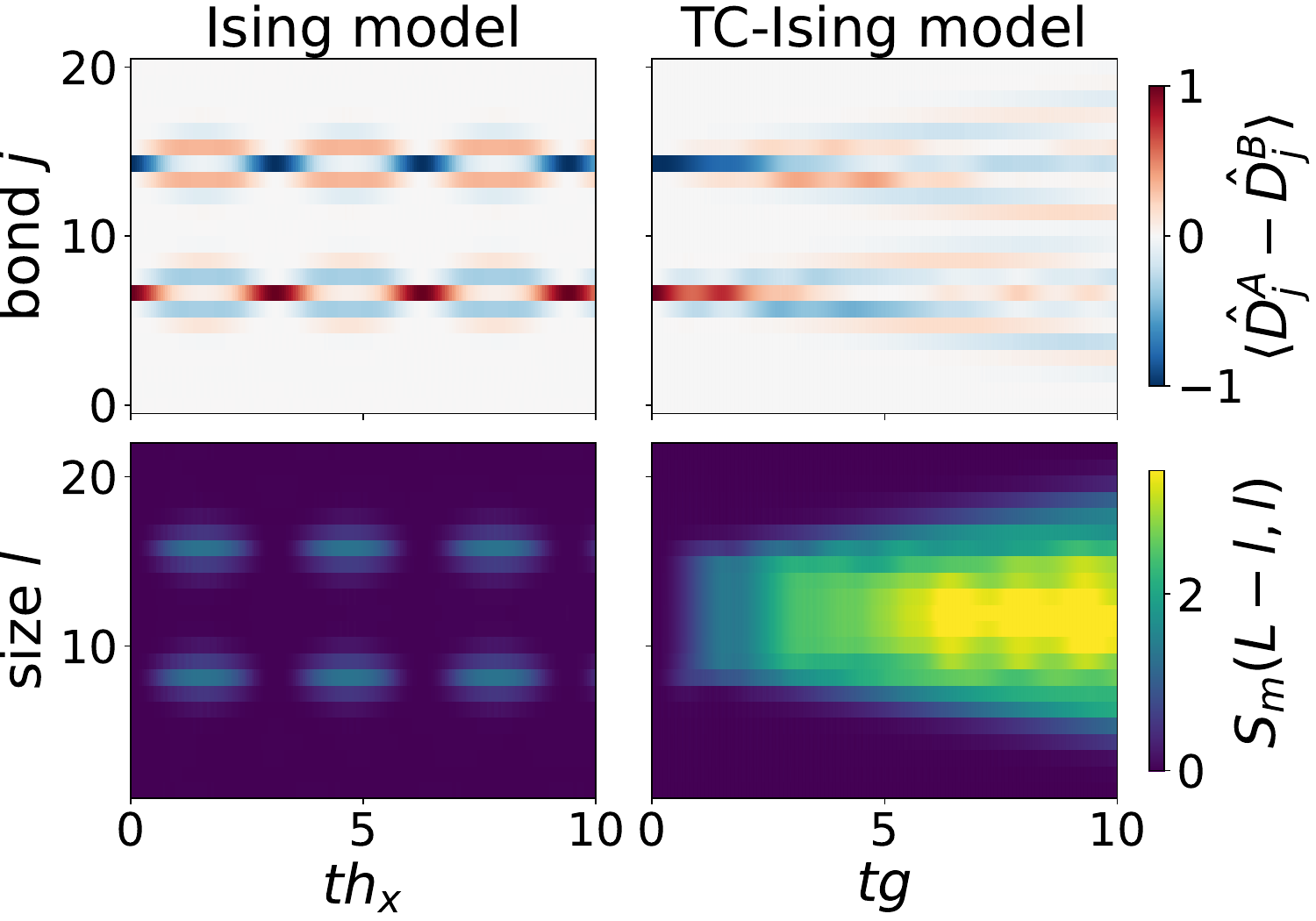}
    \put(11,40){(a)}
        \put(51,40){(b)}
    \put(11,11){\textcolor{white}{(c)}}
    \put(51,11){\textcolor{white}{(d)}}
        \end{overpic}
\caption{String dynamics comparison between Ising model (a,c) and TC-ising model (b,d) obtained with TDVP. Top panels (a,b) show the domain wall density while the bottom panel (c,d) present the bipartite mutual information $S_m(L-l,l)$. In order to have comparable dynamics, we set $h_x=0.1$ in the Ising case and $g=0.1$ in the TC-Ising one with $\delta=0.$; both cases have  $V=1$, $h_z=0.2$ and $N=23$.}
    \label{fig:string}
\end{figure}
\paragraph*{String dynamics -}In the context of confining LGT theories, an important role is played by \textit{string} excitations, namely bound states of two distant elementary excitations. In the simple case of the Ising model, two distant domain walls feel a confining potential proportional to their separation $r_0$ and the longitudinal field $h_z$. This induces an exponentially slow dynamics $(h_x/h_z)^{r_0}$ due to the underlying locality of the LGT dynamics (represented by a local transverse field $h_x$). This phenomenology is lost in the TC-Ising model where, independently of $r_0$, distant domain walls can interact via the exchange of cavity photons. We illustrate this fundamental difference in Fig. \ref{fig:string} by direct comparison of the real-time evolution of a string in the two cases of Ising model (a,c) and TC-ising model (b,d), obtained via time-dependent variational principle (TDVP) simulations \cite{supmat,White_prb2020_globalsubspaceexpansion}. The domain wall density (top panels) clearly shows that in the local Ising case strings are effectively immobile, while in the TC-Ising case strings can have a finite kinetic energy, comparable to that of single domain walls (Fig. \ref{fig:1dw_dynamics}). In particular the process which allow for the motion of arbitrary length strings $r_0$ is the following:
\begin{align}
&|..\spind\spinu\overbrace{\spinu\spind\spinu\spind\spinu\spind}^{r_0}\spind\spinu\spind..\rangle\ket{0}\rightarrow|..\spind\spinu\overline{\spind}\overbrace{\spind\spinu\spind\spinu\spind}^{r_0-1}\spind\spinu\spind..\rangle\ket{1}\rightarrow \nonumber \\ \rightarrow&|..\spind\spinu\overline{\spind}\overbrace{\spind\spinu\spind\spinu\spind\overline{\spinu}}^{r_0}\spinu\spind..\rangle\ket{0}
\end{align}
with a rate $J_s=g^2/(\delta \pm2 h_z)$ depending on whether the hopping is towards the direction where the type A domain wall is or not.

In order to confirm that the important emergent dynamical degree of freedom is the string and that single domain walls do not move independently, we calculate the mutual information (Fig \ref{fig:string}(c,d)) between two subsystem of length $l$ and $L-l$:
\begin{align}
    S_m(L-l,l)=  S(\rho_{1..l}) +S(\rho_{l+1..L}) - S(\rho_{c}) ,
\end{align}
where $S(\rho_X)$ is the von Neumann entropy of a subsystem $X$ being either the first $l$ atoms, the last $L-l$ atoms or the cavity $c$. In the case of the Ising model this just reduces to twice the bipartite entanglement entropy. The important different feature distinguishing the Ising and TC-Ising case is the presence of a \textit{plateau} already at early times ($t\lesssim 50$) when the bipartition length $l$ varies along the string ($8<l<14$). This highlights the immediate build-up of long-range correlations between the two domain walls forming the end of the string, which then can move coherently as a single object via the exchange of virtual photons.

\paragraph{Experimental blueprint. -} 
The proposed experimental implementation uses \Rb atoms trapped in optical tweezers, though the model Hamiltonian and excitation scheme described above are applicable to a range of atomic species, including alkali \cite{saffman2016quantum} and alkaline-earth-like atoms \cite{ma2022universal}.
Rydberg control is achieved via a two-photon excitation from the ground state $\ket{g} = \ket{5S_{1/2} F=2, m_F=-2}$ to the Rydberg state $\ket{r}= \ket{70S_{1/2} , J= 1/2, m_J = -1/2}$ via an intermediate state $\ket{e} = \ket{6P_{3/2} , F = 3, m_F = -3}$ \cite{bernien2017probing}. 
This transition is driven by circularly polarized lasers at wavelengths of \SI{420}{\nm} and \SI{1013}{\nm}, respectively, with an intermediate state detuning $\Delta \simeq 2\pi \times$\SI{0.5}{\GHz}. The set-up allows for a simple state preparation scheme of specific initial states~\cite{supmat}.

The one-dimensional array of atoms is coupled to a bow-tie cavity that sustains two counter-propagating unidirectional modes, of opposite circular polarization. 
This configuration balances the requirements of strong cavity coupling, compatibility with Rydberg physics, and scalability to large atom arrays. 
The cavity design provides substantial optical access, enabling hundreds of optical tweezers for single-atom control while maintaining a large distance between the atoms and the dielectric surfaces of the mirrors, thereby mitigating stray charge effects on Rydberg atoms \cite{thiele2015imaging}. 
Importantly, this design also achieves small mode volumes, ensuring the strong cavity coupling essential for the proposed implementation \cite{chen2022high}.

Assuming a cavity waist (w$_0$) of \SI{3}{\um}, a cavity length of \SI{6.9}{\cm}, a finesse of 50000 along with the measured decay rates and branching ratios of the $6P_{3/2}$ states \cite{das2024direct}, the cavity parameters are $\{g_0, \kappa, \Gamma\} = 2\pi \times \{800~\text{kHz}, 20~\text{kHz},1.35~\text{MHz} \}$. 
Here, $g_0$ represents the on-resonance atom-cavity coupling, $\kappa$ denotes the half-linewidth of the cavity, and $\Gamma$ the full linewidth of the intermediate state. 
Notably, planar ring cavities with similar w$_0/\lambda_c$ ratios have already been demonstrated for the same finesse \cite{chen2022high}.

In order to minimize off-resonant scattering from the intermediate state ($\Gamma \Omega^2/\Delta^2$) and maximize the two-photon transition strength ($g_0 \Omega /\Delta$), we propose to operate the \SI{1013}{\nm} laser at a Rabi frequency $\Omega\simeq \Delta /10 \simeq 2\pi \times 50~\text{MHz}$. The total effective atom decay, which includes direct Rydberg decay $\Gamma_r=2\pi\times 1.5\mathrm{kHz}$, is estimated as $\gamma_{at}/g= [\Gamma_r+\Gamma(\Omega/\Delta)^2]/g = 0.18$. Togheter with a cavity photon loss rate $2\kappa/g=0.5$, this justify the choice of studying the coherent dynamics up to $tg \lesssim 10$. As we discuss in detail in \cite{supmat}, simple post-selection strategies which mitigate the effect of atom decay are likely needed. The effect of photon loss is not as severe due to small cavity population during the dynamics \cite{supmat}.


\paragraph*{Conclusions -} We have studied real-time dynamics of low energy excitations in a Tavis-Cumming-Ising model. This highlights the non-trivial interplay between local and non-local degree of freedoms, contrasted with usual locally interacting models. Real time and space resolved dynamics in particular is key to distinguish the degree of ``locality" of the emergent composite excitations in the confined phase. We further expect new interesting physics arising also in the deconfined phase, where non-local interactions dominate from the start giving rise to an otherwise forbidden continuous $U(1)$ SSB in one dimension \cite{Li_2021pra}. All these phenomena naturally arise in a simple combination of two well-developed quantum simulation platforms, Rydberg atom arrays \cite{bernien2017probing,Browaeys2020} and atomic cavity QED \cite{Lonard_nature2017,yan_2023_Stamper_superradiant}. Nonetheless, the physical ideas explored in such simple setting can be translated to other cavity embedded quantum many-body systems \cite{Schlawin_2022apr_cavityrev,GarciaVidal_2021science_cavityrev,Bloch_2022nat_cavityrev,Mivehvar2021}, where local and non-local emergent quasi-particles coexist.

\paragraph{Acknowledgments -}We thank T. Chanda, D. Chang, M. Collura, M. Oehlgrien and F. Scazza for discussions. MPS based simulations have been carried out with the help of the libraries ITensors.jl \cite{ITensor} and TenNetLib.jl \cite{tennetlib}. ED and quantum trajectory base simulations have been carried out with QuTip \cite{Qutip}. 
M.\,D. was partly supported by the QUANTERA DYNAMITE PCI2022-132919. M.\,D. was supported by the EU-Flagship programme Pasquans2, by the PNRR MUR project PE0000023-NQSTI, the PRIN programme (project CoQuS), and the ERC Consolidator grant WaveNets. H.\,B.\,X. was supported by the MIUR programme FARE (MEPH).
\bibliography{main_v2_suppmat}

\clearpage
\onecolumngrid

\begin{center}
    {\large\bfseries Supplementary Material for
``Local vs non-local dynamics in cavity-coupled Rydberg atom arrays''
    } \\[0.5cm]
    {Zeno Bacciconi, Hernan B. Xavier, Matteo Marinelli, Devendra Singh Bhakuni, Marcello Dalmonte
   }
\end{center}
\setcounter{equation}{0}
\setcounter{figure}{0}
\renewcommand{\theequation}{S\arabic{equation}}
\renewcommand{\thefigure}{S\arabic{figure}}
\renewcommand{\thesection}{\arabic{section}}


\section{Derivation of the Hamiltonian}\label{app:H_der}
\begin{figure}[h!]
    \centering
    \includegraphics[width=4cm]{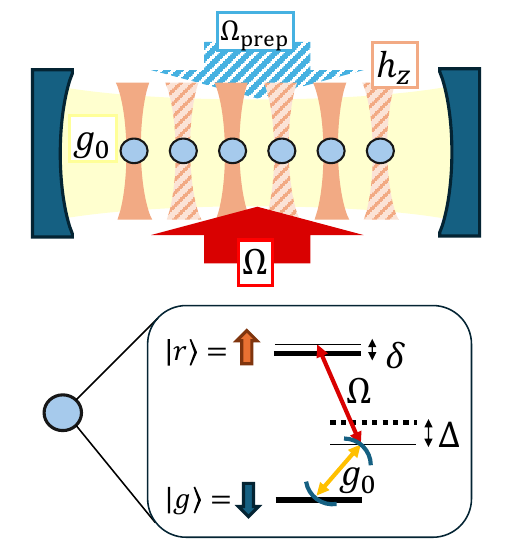}
    \caption{Light-atom setup. A linear Rydberg atom array is put inside a cavity. A two-photon transition is used to drive atoms from the ground state to the Rydberg state.}
    \label{fig:setup}
\end{figure}

In this section, we detail the derivation that motivates model Hamiltonian, Eq. (\textcolor{blue}{1}), studied in the main text. The system under consideration involves an atomic array put inside a cavity, as shown in Fig. \ref{fig:setup}, positioned in the maxima of the cavity field. Atoms are excited from the ground state $\ket{g}$, to a Rydberg state $\ket{r}$, via an intermediate state $\ket{e}$. The ground-to-intermediate state transition is driven by an off-resonant cavity mode, characterized by a coupling strength $g_0$, assumed equal for all atoms, and detuning $\Delta$. The subsequent transition from the intermediate state to the Rydberg state is mediated by a classical laser field with a Rabi frequency $\Omega$ again assumed to be  equal for all atoms. The combined two-photon transition $\ket{g}\leftrightarrow\ket{r}$ has detuning $\delta_0$. In the rotating frame, the model Hamiltonian is expressed as 
$\hat{H}_{rf}=\hat{H}_{0}+\hat{H}_{1}$, where
\begin{align}\label{eq:Hrf}
    \hat{H}_{0}&=
    \delta_0\hat{a}^\dagger\hat{a}
    +(\Delta+\delta_0)\sum_j\hat{\sigma}_{j}^{ee}
    -\sum_jw_j\sigma_j^{gg}
    +V_\text{NN}\sum_{j}\hat{\sigma}_{j}^{rr}\hat{\sigma}_{j+1}^{rr},
    \nonumber\\
    \hat{H}_{1}&=g_0\sum_j(\hat{a}\hat{\sigma}_{j}^{eg}+\hat{a}^\dagger\hat{\sigma}_{j}^{ge})
    +\Omega\sum_j(\hat{\sigma}_{j}^{re}+\hat{\sigma}_{j}^{er}).
\end{align}
The Rydberg-Rydberg interaction is here truncated for simplicity.The canonical boson operator $\hat{a}$ annihilates cavity photons, and projectors $\hat{\sigma}^{\alpha\beta}=\ket{\alpha}\bra{\beta}$ with $\alpha=g$, $e$, and $r$ describe atomic operators. 
Additionally, we introduce a site-dependent detuning
$w_j$ induced by AC Stark shifts from the optical tweezers,
along with the nearest-neighbor interaction
$V_\text{NN}$ for atoms in the Rydberg state.

In the far off-resonant limit $\Delta\gg1$, the intermediate state is barely occupied. We adiabatically eliminate it by means of a Schrieffer-Wolff transformation defined as $\hat{H}'=e^{\hat{S}}\hat{H}_{rf}e^{-\hat{S}}$. We consider a generator $\hat{S}=\hat{S}_c+\hat{S}_l$, where $\hat{S}_c$ and $\hat{S}_l$ represent contributions from the cavity and the laser.  The generators are chosen so that their commutator with $\hat{H}_0$ produce $\comm{\hat{S}}{\hat{H}_0}=-\hat{H}_1$.
Explicitly, they read:
\begin{align}
    \hat{S}_c&=
    g_0\sum_j
    \big(\Delta+w_j\big)^{-1}
    (\hat{a}\hat{\sigma}_j^{eg}-\hat{a}^\dagger\hat{\sigma}_j^{ge}),
    \nonumber\\
    \hat{S}_l&=
    \Omega\sum_j\Big(\Delta+\delta_0
    -V_\text{NN}\hat{\sigma}_{j+1}^{rr}
    -V_\text{NN}\hat{\sigma}_{j-1}^{rr}\Big)^{-1}
    (\hat{\sigma}_j^{er}-\hat{\sigma}_j^{re}).
\end{align}
The laser term includes a site-dependent denominator operator $\hat{D}_j=\Delta+\delta_0-V_\text{NN}\hat{\sigma}_{j+1}^{rr}-V_\text{NN}\hat{\sigma}_{j-1}^{rr}$. This operator is diagonal in the  eigenbasis of $\hat{H}_0$, and accounts for the local atomic configuration. The commutators of the generators with $\hat{H}_0$ are given by:
\begin{equation}
    \comm{\hat S_c}{\hat{H}_0}=
    -g_0\sum_j(\hat{a}\hat{\sigma}_{j}^{eg}+\hat{a}^\dagger\hat{\sigma}_{j}^{ge}),
    \qquad
    \comm{\hat S_l}{\hat{H}_0}=
    -\Omega\sum_j(\hat{\sigma}_j^{er}+\hat{\sigma}_j^{re}).
\end{equation}
With this choice the condition $\comm{\hat{S}}{\hat{H}_0}=-\hat{H}_1$ is met identically, and the effective Hamiltonian acquires the form:
\begin{equation}
    \hat{H}_{eff}=
    \hat{P}\hat{H}_0\hat{P}
    +\frac12\hat{P}\comm{S}{\hat{H}_1}\hat{P}
    +\cdots,
\end{equation}
where $\hat{P}=\prod_j(1-\hat\sigma_j^{ee})$ is the projector onto the subspace without atoms in the excited state. The second-order corrections to the Hamiltonian are obtained from the commutator $\comm{S}{\hat{H}_1}$. We find that the contribution coming from $\hat S_c$ reads
\begin{equation}
    \hat{P}\comm{\hat S_c}{\hat{H}_1}\hat{P}=
    -\frac{2 g_0^2}{\Delta}\hat{a}^\dagger\hat{a}\sum_j\hat{\sigma}_j^{gg}
    -\frac{g_0\Omega}{\Delta}\sum_j(\hat{a}\hat{\sigma}_{j}^{rg}+\hat{a}^\dagger\hat{\sigma}_{j}^{gr})
    +\cdots,
\end{equation}
where we project the result to the manifold of interest, 
and approximate $\frac{1}{\Delta+w_j}\approx\frac{1}{\Delta}+\cdots$, as we concentrate on the regime where $\Delta$ is the dominant energy scale, i.e., $\Delta\gg |w_j|$.
Likewise, the contribution from $\hat S_l$ is obtained from the commutator:
\begin{align}
    \hat{P}\comm{\hat S_l}{\hat{H}_1}\hat{P}
    &=-\frac{g_0\Omega}{\Delta}\sum_j(\hat{a}\hat{\sigma}_{j}^{rg}+\hat{a}^\dagger\hat{\sigma}_{j}^{gr})
    -\frac{2\Omega^2}{\Delta}\sum_j\hat\sigma_j^{rr}
    +\cdots.
\end{align}
Note that terms arising from the commutation of $\hat D_j^{-1}$ vanish under projection, and the final expression follows from the large-$\Delta$ approximation $\hat{D}_j\approx\Delta$ 
. Within this limit, the second-order effective Hamiltonian simplifies to
\begin{align}\label{eq:Heff}
    &\hat{H}_{eff}=
    \delta\hat{a}^\dagger\hat{a}
    +\frac12\sum_jh_j\hat{\sigma}_j^{z}
    +\frac{V_\text{NN}}{4}\sum_{j}\hat{\sigma}_{j}^{z}\hat{\sigma}_{j+1}^{z}
    +g\sum_j(\hat{a}\hat{\sigma}_{j}^{+}+\hat{a}^\dagger\hat{\sigma}_{j}^{-})
    +\lambda\hat{a}^\dagger\hat{a}\sum_j\hat{\sigma}_j^{z}.
\end{align}
In passing to the spin notation we ignore additive constants and boundary terms. The mapping is achieved through the identification  
$\hat\sigma_j^+=\hat\sigma_j^{rg}$, with 
$\hat\sigma^z_j=\hat\sigma^{rr}_j-\hat\sigma^{gg}_j$.
The effective couplings are expressed as follows: 
$\delta=\delta_0-Ng_0^2/2\Delta$,
$h_j=w_j+V_\text{NN}-\Omega^2/\Delta$, 
$g=-g_0\Omega/\Delta$, and
$\lambda=g_0^2/2\Delta$.

\section{State preparation}

We here focus on the state preparation protocol. To realize the physics discussed in this work, the atoms are initialized in the classical states described in the main text through a combination of global excitations and local addressing beams that detune specific atoms from resonance. 
All atoms are initially trapped and prepared in the ground state $\ket{g}$.
To prepare atoms in the Rydberg state $\ket{r}$, the optical tweezers trapping these atoms are turned off, and a two-photon excitation scheme is applied \cite{bernien2017probing}. 
This scheme uses two laser beams that globally address the array: a \SI{1013}{\nm} beam (also used during subsequent coherent dynamics) and a \SI{420}{\nm} beam external to the cavity (marked as $\Omega_{\mathrm{prep}}$ in Fig. \ref{fig:setup}). 
The relative frequency of the lasers is chosen such that the cavity is detuned by several tens of MHz from the \SI{420}{\nm} beam, avoiding the risk of populating the cavity mode with undesired photons.
Atoms still illuminated by the optical tweezers are detuned by tens of MHz from the two-photon excitation, ensuring they remain in the ground state. This selective addressing allows the realization of the desired classical initial state.
At the end of the state preparation, the external \SI{420}{\nm} beam is turned off, along with all optical tweezers, and the frequency of the \SI{1013}{\nm} beam is adjusted to establish a two-photon resonance with the optical cavity.
\section{MPS numerical details}
We use an MPS ansatz for hybrid cavity-matter systems which has been already used for example in \cite{HalatiKollath_prl2020_mpscavity,bacciconi2023topological}. The cavity site is positioned on one end of the MPS, with a truncated Hilbert space $N_{ph}$. For ground state simulations we use $N_{ph}=200$ and a maximum bond dimension $\chi=200$ allowing us to keep the truncation error below $10^{-8}$ in the $U(1)$ SSB phase which is the more entangled of the two. For TDVP dynamics done in the confined phase we use $N_{ph}=20$ and $\chi=100$. Here we find a poor performance of a simple 2-site TDVP update scheme when starting from the classical states describe in the main text. We thus make use of global subspace expansion updates as introduced in \cite{White_prb2020_globalsubspaceexpansion} and available in the library TenNetLib.jl \cite{tennetlib}. Then, in order to calculate the mutual information we perform a series of site swaps of the cavity site at specific times during the dynamics.
\section{Ground state details}\label{app:pd}
\begin{figure}
    \centering
    \includegraphics[width=0.9\linewidth]{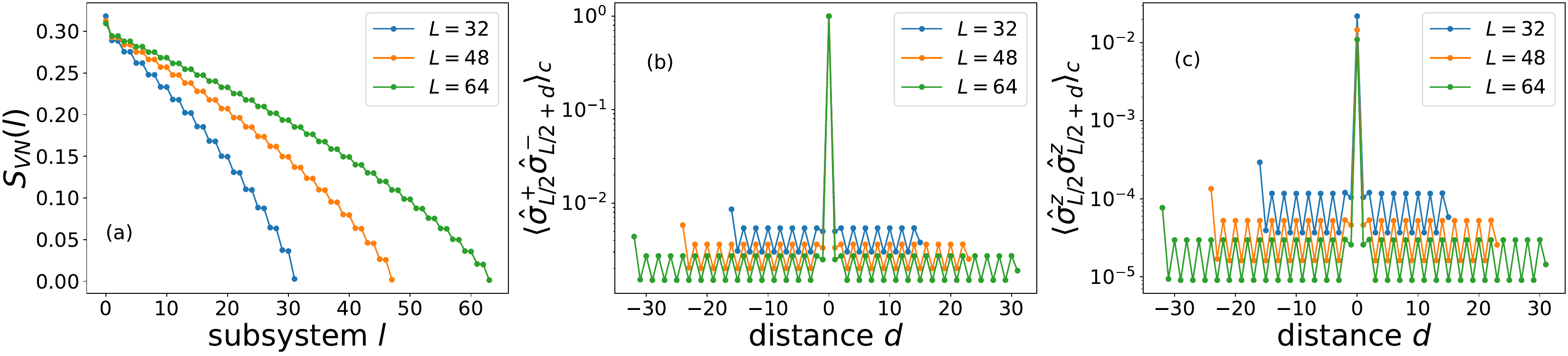}
    \caption{Ground state properties of the confined phase obtained with DMRG. (a) Von Neuman entanglement entropy of a bipartition of $l_{at}$ atoms plus the cavity ($l_{at}=0$ is the cavity-atom entanglement). (b-c) Connected correlation functions showing the infinite range nature of correlations in the phase. In all panels we fix $G=g\sqrt{N}=4.5$, $h_z=0.5$, $\delta=1$ and $\lambda=0.$.   }
    \label{fig:ground_state}
\end{figure}

We here report a more detailed study on the ground state properties of the confined phase, whose low-energy dynamics is studied in the main text. In order to access large system size we use DMRG simulations with a mixed MPS ansatz where the cavity sits at one side of the atom chain. In figure \ref{fig:ground_state}(a) we show the bipartite entanglement entropy of the system as a function of bipartition size. At $l_{at}=0$ this correspond to the overall cavity-atom entanglement entropy, which remains finite as the system size is increased. This contribution comes from virtual ground state mixing of meson-polaritons discussed in the main text. Then in figure \ref{fig:ground_state}(b-c) we show two connected atom-atom correlation functions, respecitely in plane and zz. They both show feature of infinite range correlations which become less relevant in the $N\to \infty $ limit. In particular the former decays as $1/N$ while the second one as $1/N^2$. 
\section{Effect of $r^{-6}$ range interactions}
\begin{figure}[h!]
    \centering
    \includegraphics[width=0.7\linewidth]{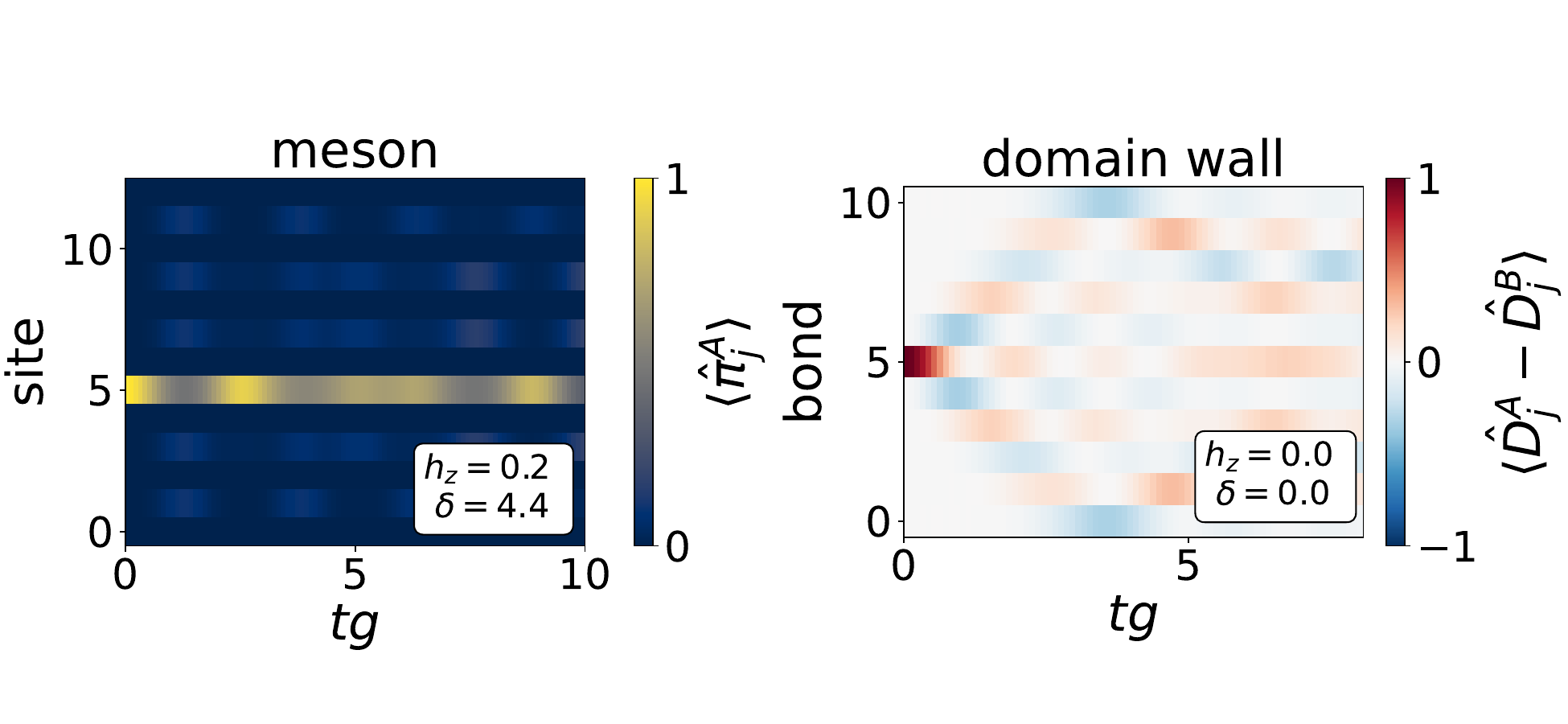}
    \caption{Effect of $r^{-6}$ interaction on meson and domain wall dynamics. (a) Meson dynamics at $h_z=0.2$ and $\delta=4V+2h_z=4.4$, slightly missing the resonance condition because of a renormalized meson energy $E_\pi \simeq 4.36$. (b) Domain wall dynamics at $\delta=h_z=0$ left unchanged by the long-range part as the domain wall number does not change during the dynamics. }
    \label{fig:long_range}
\end{figure}

In this section we discuss the effect of the weak long-range tail of the Rydberg-Rydberg interaction. In this case the Hamiltonian reads:
\begin{align}\label{eq:H_lr}
    &\hat{H}_{lr}= \delta \hat{a}^\dagger \hat{a} +  h_z\sum_{j}(-1)^j \hat{\sigma}^z_j + \frac{1}{2}\sum_{i\neq j} \frac{V}{|i-j|^6}\hat{\sigma}^z_i\hat{\sigma}^z_{j} + g\sum_j\left(\hat{a}\hat{\sigma}^+_j + \hat{a}^\dagger \hat{\sigma}_j^-\right)+ \lambda\; \hat{a}^\dagger \hat{a}\sum_{j}\hat{\sigma}_j^z\;.
\end{align}
Already the next-nearest neighbour term ($i=j\pm2$) gives contributions on the order of $V/2^6 \sim 0.015 V$, smaller than the values of $g$ studied in the manuscript. In Figure \ref{fig:long_range} we quantitatively asses that the dynamics of domain-walls and meson-polaritons is not affected by long-range interactions. In particular we can understand from the perturbation theory arguments depicted in the main text that the propagation of single domain walls or the interaction between distant ones in the case of strings, is not affected by the $r^{-6}$ tail. This just modify the classical energy of a domain wall to be $2V-4\frac{V}{2^6}+6\frac{V}{2^{12}}+...$ irrespectively of its position or type (A or B) and thus do not affect string and domain wall dynamics. Instead the meson-polariton dynamics is slighly changed as the resonance condition depends on the meson energy. In particular the meson A energy in presence of $r^{-6}$ interactions becomes $E_{\pi_A}=4V+2h_z-4\frac{V}{2^6}-4\frac{V}{2^{12}}+...\simeq 4.34$. 
\section{Cavity losses }
\begin{figure}[h!]
    \centering
    \begin{overpic}[width=0.6\linewidth]{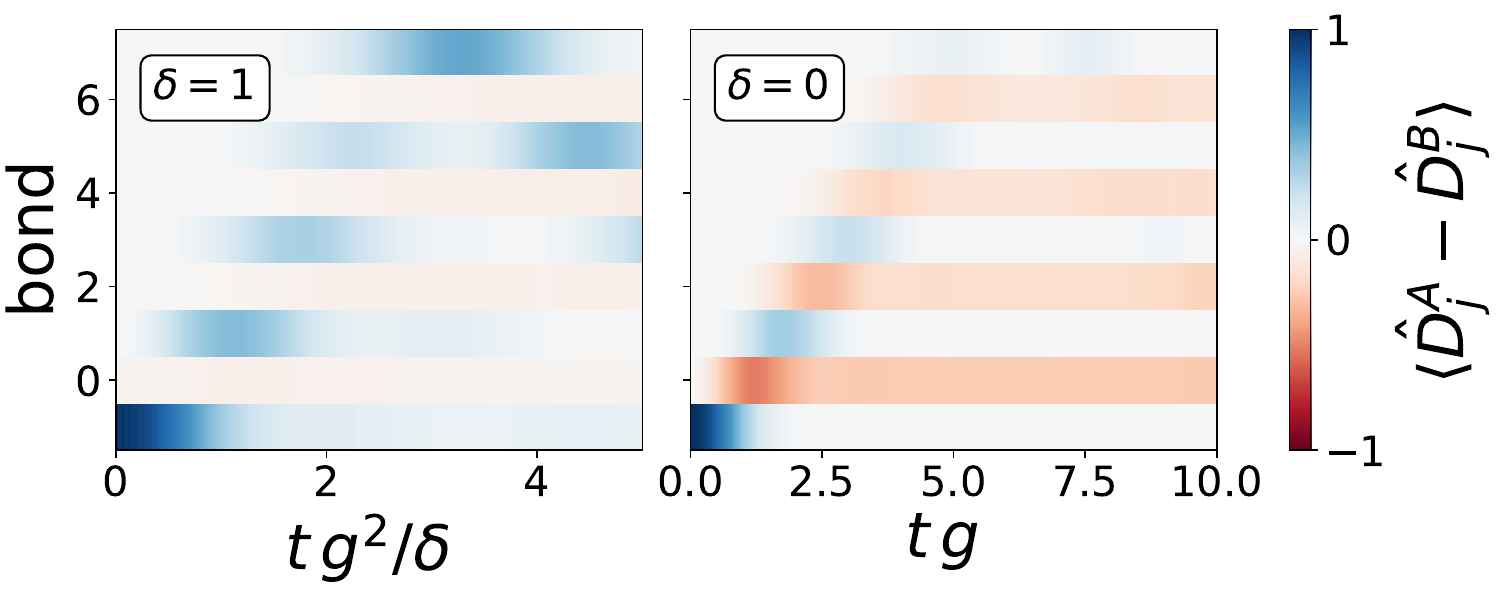}
    \put(35,12){(a)}
    \put(75,12){(b)}
    \end{overpic}
    \caption{Effect of finite cavity photon losses $\kappa=g/2=0.05$ on domain wall dynamics for (a) off-resonant $\delta=1$ and (b) resonant $\delta=0$ cavity detunings. The initial state is prepared with a boundary domain wall of type A. $h_z=0$ and $N=8$.}
    \label{fig:losses}
\end{figure}
The loss mechanism expected to be dominant in the set-up considered in this work is the loss of cavity photons, with a rate $\kappa\lesssim g$. We consider explicitly their effect on the single domain-wall dynamics via averaging quantum trajectories. In Fig. \ref{fig:losses} we show two scenarios for the cavity detuning $\delta=0$ (a) and $\delta=1$ (b) in a case of of weak-intermediate cavity losses $\kappa=g/2=0.05$. In particular we look at the dynamics of a type A domain wall initialized at the boundary of the atom array. This does not correspond to two actual initial rydberg atoms close to each other, but just the addition of a further boundary longitudinal field mimicking the presence of another rydberg atom. Note indeed that the bond at which the domain wall is $j=-1$, before the first atom.

The photon loss projects the system into states with a domain wall of type B instead of type A, giving the overall effect of freezing the dynamics as $J_B\ll J_A,g$. The actual timescale at which the domain wall of type A \textit{decays} into a type B depend on the specific scenario.  More in details we can see that the effect of photon losses is stronger on the $\delta=0$ case where cavity photons are coherently generated during the dynamics and the lifetime for the domain wall propagation is on the order of $\kappa^{-1}$. Instead in the $\delta=1$ case the cavity is only virtually populated $\langle \hat{n}_{ph}\rangle \sim \frac{g^2}{\delta^2}$, thus inducing a longer lifetime for domain walls of type A of order $\kappa \frac{g^2}{\delta^2} $.

\section{Atom decay and post-selection}
\begin{figure}[h!]
\centering
    \begin{overpic}[width=0.6\linewidth]{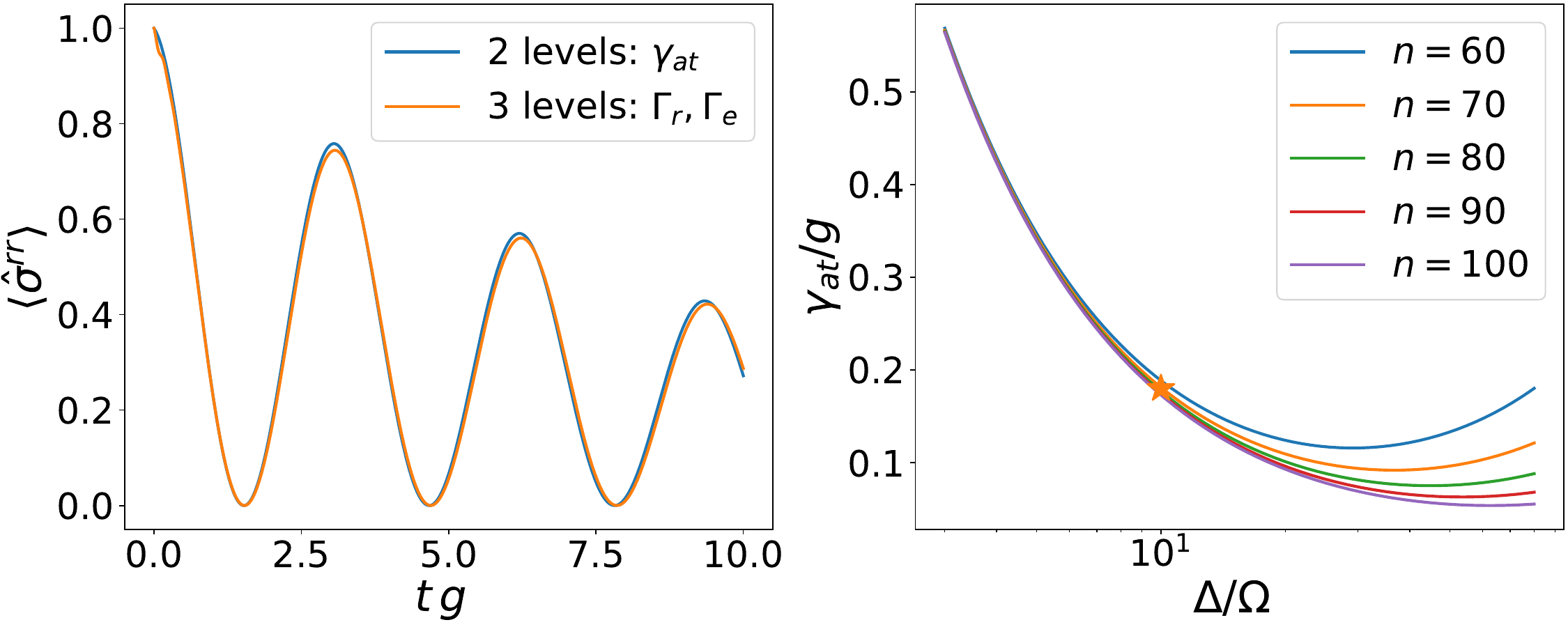}
    \put(12,36){(a)}
    \put(61,36){(b)}
    \end{overpic}

    \caption{(a) Decay of cavity-induced Rabi oscillations between the rydberg state and ground state at $g_0=2\pi\times0.8\,\mathrm{MHz}$, $\Omega=2\pi\times50\,\mathrm{MHz}$ and $\Delta=2\pi\times500\,\mathrm{MHz}$ ($g=g_0\Omega/\Delta$) in the 3-level atom scheme (orange) and in the effective 2-level atom scheme (blu). In the three level atom both rydberg decay $\Gamma_r=2\pi \times 1\,\mathrm{kHz}$ and intermediate state decay $\Gamma_e=2\pi \times1.35 \,\mathrm{MHz}$ are included, well accounted for by an effective two-level atom decay of $\gamma_{at}=\Gamma_r+\Gamma_e (\Omega/\Delta)^2$. (b) Effective atom decay in the two-level atom $\gamma_{at}/g$ as a function of intermediate state detuning $\Delta/\Omega$ for different Rydberg principal quantum number $n$, assuming a rydberg decay $\Gamma_r^n = \Gamma_r^{70} \left(\frac{70}{n}\right)^3 $ with $\Gamma_r^{70} =2\pi \times 1 \,\mathrm{kHz}$}
    \label{fig:3l_loss}
\end{figure}

We here investigate the role of atom decay. The Rydberg state considered in this work has an intrinsic lifetime of $\tau \sim 150\,\mathrm{\mu s}$ leading to a decay rate to the ground state of $\Gamma_r=2\pi\times 1\,\mathrm{kHz}$. Virtual population of the intermediate state leads to another decay channel for the rydberg state to the ground state of $\Gamma_{eff}=\Gamma_e (\Omega/\Delta)^2 \sim 2\pi\times 13 \,\mathrm{kHz}$ leading to a total decay rate of the two-level atom ($\ket{r}$ and $\ket{g}$) of $\gamma_{at}=\Gamma_r+\Gamma_{eff}=2\pi \times14 \,\mathrm{kHz}$. This is explicitly shown in Fig. \ref{fig:3l_loss}(a) where decay of Rabi-oscillations $\ket{r}\to\ket{g}$ in the two level atom model is compared to the full three level atom without integrating out the intermediate state. In Fig. \ref{fig:3l_loss}(b) we further show the dependence of $\gamma_{at}/g$ on $\Omega/\Delta$ for different choice of Rydberg atom principal numbers $n$.

\begin{figure}[h!]
\centering
    \begin{overpic}[width=0.9\linewidth]{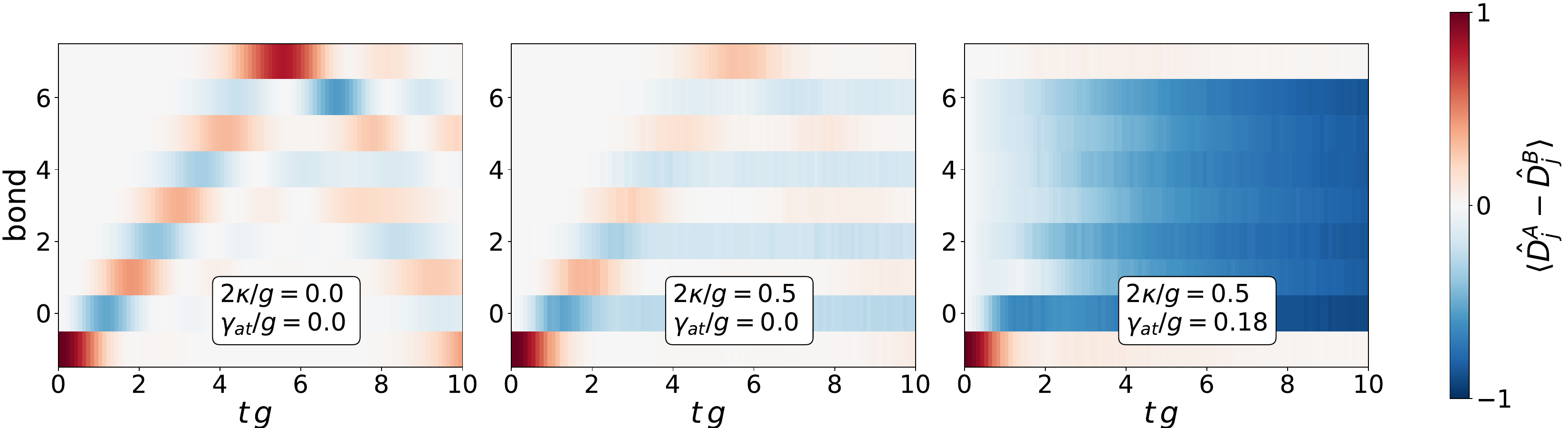}
    \put(4,22){(a)}
    \put(33,22){(b)}
    \put(63,22){(c)}
    \end{overpic}
    \begin{overpic}[width=0.6\linewidth]{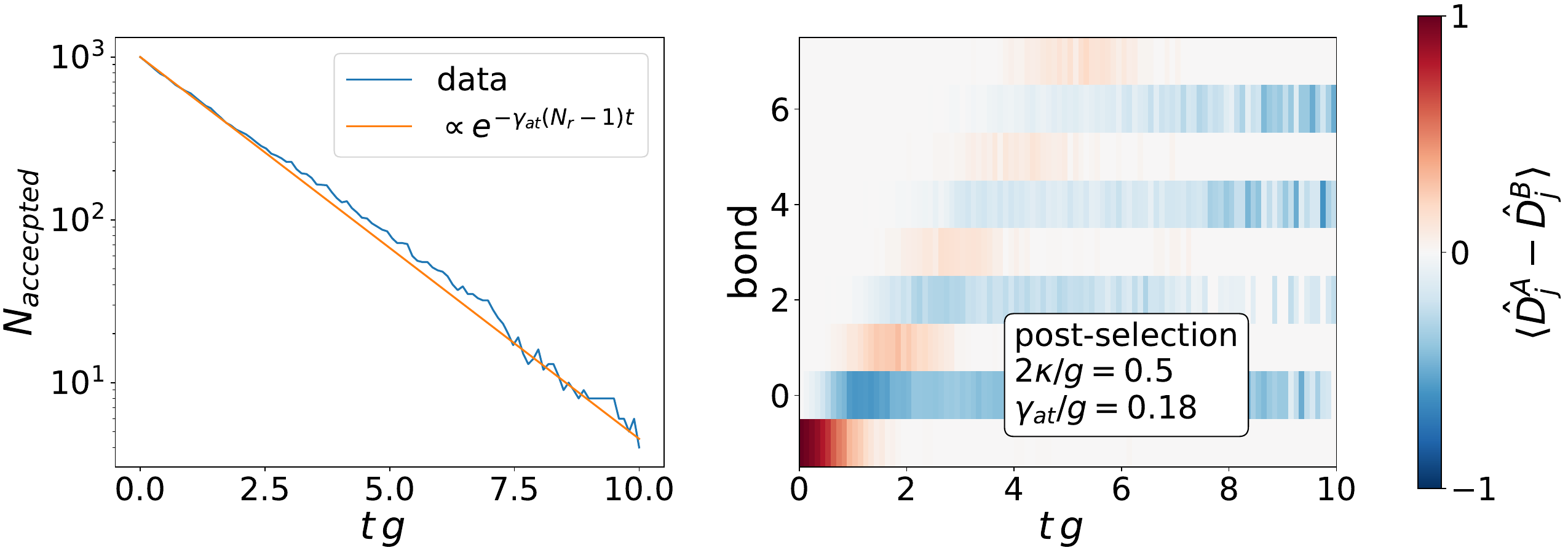}
    \put(8,8){(d)}
    \put(53,29){(e)}
    \end{overpic}
    \caption{Effect of atom decay and cavity photon loss on the dynamics of a single domain wall of type A at $g=0.1 \,V$, $\delta=0.$ and $V=1$. Top row (a,b,c) shows the doain wall density obtained by averaging 1000 trajectories with respectively (a) no losses $2\kappa=0.$,$\gamma_{at}=0.$ (b) only photon loss $2\kappa/g=0.5$,$\gamma_{at}/g=0.$ (c) both photon and atom decay $2\kappa/g=0.5$,$\gamma_{at}/g=0.18$. Bottom row (d,e) shows post-selection results (as described in the text) with both photon loss and atom decay $2\kappa/g=0.5$,$\gamma_{at}/g=0.18$. In particular (d) shows the number of accepted atom array measurements ($N_{accepted}$) and (e) shows the domain wall density obtained from these snapshots.}
    \label{fig:atom_decay_dw}
\end{figure}

We now explore the effect that the atom decay has on the dynamics under consideration. In Fig. \ref{fig:atom_decay_dw} we show different scenario to exemplify of the dynamics of a single domain wall is affected\footnote{Here we use further longitudinal fields on the boundary atoms $h_z^0=V$ and $h_z^{N-1}=V$ which mimick the presence of two further atoms fixed in the rydberg state at sites $j=-1$ and $j=N$, thus allowing a bond at $j=-1$ and $j=N-1$}.
In particular in (a) we show the unitary dynamics with no losses, in (b) the dynamics taking into account only photon loss, and in (c) the dynamics with both photon loss and atom decay. Each Rydberg decay from an the antiferromagnetic state correspond to the creation of a pair of type B domain walls, a process which at the level of unitary dynamics is strongly suppressed due to the high energy cost of a domain wall $2V\gg g$. This allow for a simple post-selection strategy on the measured atom array configurations (snapshots in the $\sigma^z$ basis) consisting in just discarding measurements which show more domain wall states than the initially prepeared state. The result of this procedure done by sampling 1000 quantum trajectories is shown in the bottom row of Fig. \ref{fig:atom_decay_dw}. In particular panel (d) shows the number of accepted snapshots as a function of time and panel (e) the resulting expectation value averaged over the accepted snapshots. For comparison, panel (c) is produced with the same 1000 trajectories but no post-selection on the snapshots. The probability of accepting a trajectory is governed by $e^{-\gamma_{at} (N_r-1)t}$ with $N_r$ the number of initial Rydberg and not $e^{-\gamma_{at} N_rt}$ because this procedure can only correct the Rydberg decay when the atom is not part of a domain wall. If the atom is part of a domain wall, its decay will not increase the total number of domain walls but just turn a domain wall of type A into type B. The number $N_r-1$ is the actual average number of atoms in the Rydberg state during the coherent evolution.

\end{document}